\documentclass[12pt,hidelinks]{article}
\usepackage[body={17.5cm, 22cm},right=2cm]{geometry}
\usepackage{color}
\usepackage{amssymb,graphicx}
\usepackage{amsmath}
\usepackage{epsfig}
\usepackage[bookmarks=true,pdfborder={0 0 0}]{hyperref}

\newcommand{\be}{\begin{equation}}
\newcommand{\ee}{\end{equation}}
\newcommand{\bea}{\begin{eqnarray}}
\newcommand{\eea}{\end{eqnarray}}
\newcommand{\bg}{\begin{gather}}
\newcommand{\eg}{\end{gather}}
\newcommand{\bseq}{\begin{subequations}}
\newcommand{\eseq}{\end{subequations}}

\begin{document}
\setcounter{page}{0}
\thispagestyle{empty}

\parskip 3pt

\font\mini=cmr10 at 2pt

\begin{titlepage}
%
~\vspace{1cm}
\begin{center}

{\Huge \sc Mini-Split}



\vspace{1.2cm}

{\large \bf
Asimina Arvanitaki$^{a}$, Nathaniel Craig$^{b}$,
Savas Dimopoulos$^{a}$,  \\[5pt] and Giovanni Villadoro$^{c}$}
\\
\vspace{.6cm}
{\normalsize { \sl $^{a}$ Stanford Institute for Theoretical Physics,\\
Stanford University, Stanford, CA 94305 USA}}

\vspace{.3cm}
{\normalsize { \sl $^{b}$ 
Rutgers University, Piscataway, NJ 08854 USA \\
{\it and} \\
Institute for Advanced Study, Princeton, NJ 08540 USA }}

\vspace{.3cm}
{\normalsize { \sl $^{c}$ SLAC, Stanford University \\
2575 Sand Hill Rd., Menlo Park, CA 94025 USA}}

\end{center}
\vspace{.8cm}
\begin{abstract}
The lack of evidence for new physics beyond the standard model at the LHC points to a paucity of new particles near the weak scale. This suggests that the weak scale is tuned and that supersymmetry, if present at all, is realized at higher energies. The measured Higgs mass constrains the  scalar sparticles to be below $10^5$ TeV, while gauge coupling unification favors Higgsinos below 100 TeV. Nevertheless, in many models gaugino masses are suppressed and remain within reach of the LHC. Tuning the weak scale and the renormalization group evolution of the scalar masses constrain Split model building. Due to the small gaugino masses, either the squarks or the up-higgs often run tachyonic; in the latter case, successful electroweak breaking requires heavy higgsinos near the scalar sparticles. We discuss the consequences of tuning the weak scale and the phenomenology of several models of Split supersymmetry including anomaly mediation, $U(1)_{B-L}$ mediation, and Split gauge mediation. 
\end{abstract}

\end{titlepage}

\tableofcontents

\section{The Missing Superpartner Problem}
\label{intro}

\begin{figure}[t!] 
 \begin{center}
 \includegraphics[width=0.4\textwidth]{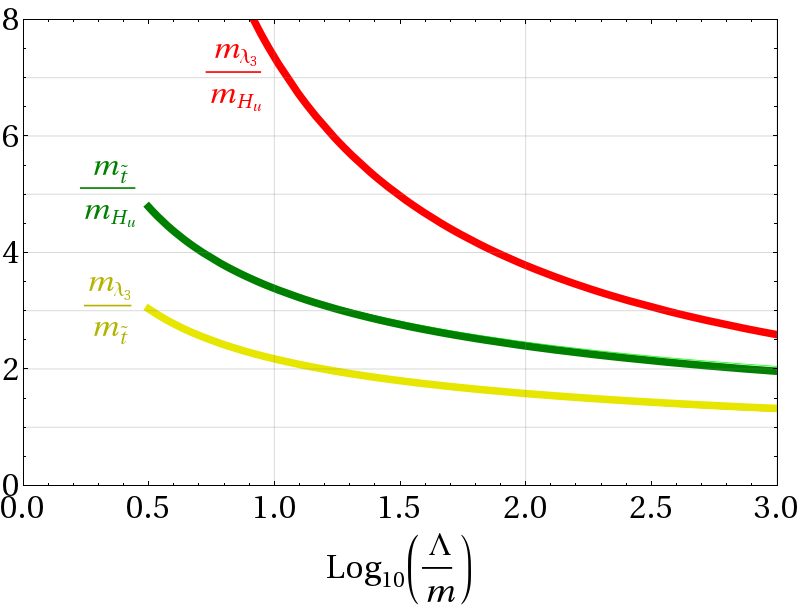} 
 \caption{ Natural ratios of gluino, stop, and up-Higgs soft masses as a function of the one-loop RG running between the soft scale $m$ and the messenger scale $\Lambda$, assuming the scalar soft masses are generated radiatively from the gluino mass.}
 \label{fig:gluino}
 \end{center}
\end{figure}

\begin{figure}[h!] 
 \begin{center}
 \includegraphics[width=.9\textwidth]{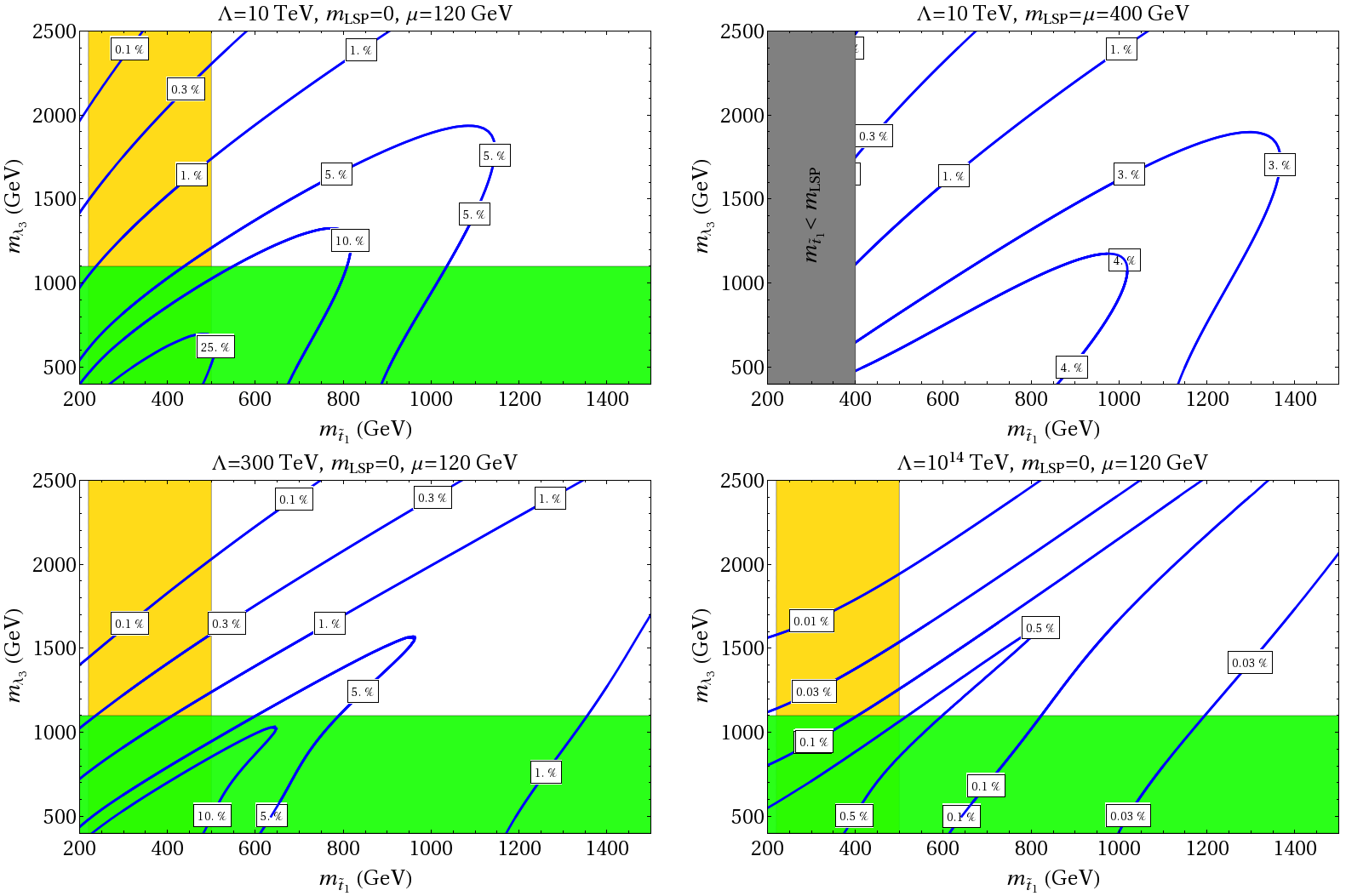} 
 \caption{The fine-tuning in the gluino--lightest stop ($\tilde t_1$) mass plane for various values of $\Lambda$ and the LSP mass. The green and yellow shaded areas correspond to the excluded regions from direct gluino and stop production searches at the LHC \cite{naturalsusybounds}, respectively. }
 \label{fig:tuning}
 \end{center}
\end{figure}

\begin{figure}[t!] 
 \begin{center}
 \includegraphics[width=0.6\textwidth]{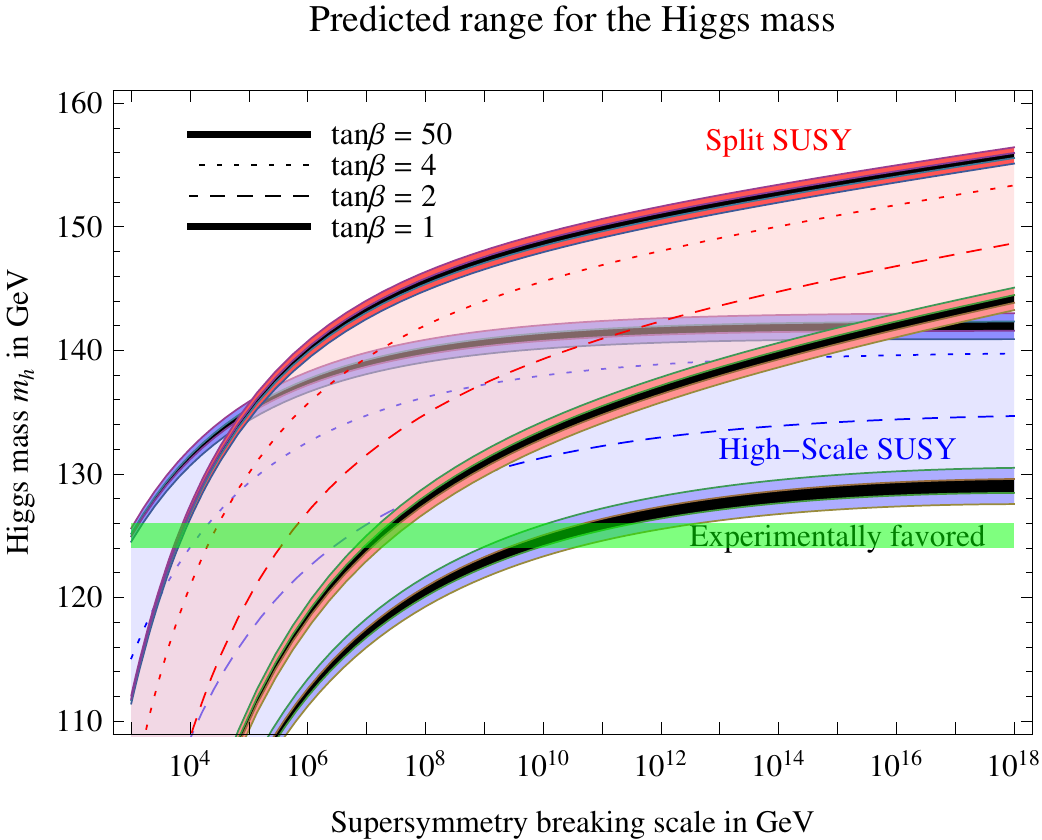} 
 \caption{The Higgs mass prediction as a function of the scalar mass scale in Split and High-Scale Supersymmetry for different values of $\tan \beta$, taken from \cite{strumia}.}
 \label{fig:strumia}
 \end{center}
\end{figure}

Particle physics is at a crossroad. On one hand is the road of naturalness, leading to weak-scale supersymmetry (SUSY), or perhaps to TeV-scale gravity or compositeness. On the other is the less travelled road of the multiverse, leading to Split supersymmetry or the standard model. We live in exciting times when the LHC may resolve this dichotomy in this decade. 

The confirmation of the supersymmetric prediction of gauge coupling unification \cite{drw81,dg81} in the early '90s by LEP and SLC gave a tremendous boost to naturalness and weak-scale SUSY, and led to the expectation of a major discovery by LEP2. Unfortunately this was not to be. By the late nineties the absence of the Higgs  showed that simple versions of supersymmetric theories were  tuned at the $\leq10\%$ level. This ``little hierarchy problem" left open the possibility that naturalness was realized in more involved theories.

	The LHC has significantly changed the prospects for naturalness. The value of the Higgs mass in the context of the MSSM points to a top squark $\gtrsim$~few~TeV, which typically implies a tuning below the percent level. This may be cured at the expense of going beyond the MSSM, for example, to the NMSSM or theories with extra gauge groups. The direct superpartner limits, however, which are now beginning to exceed $\sim$1.5 TeV for all colored sparticles \cite{atlassusybound} except the stop and sbottom, pose a serious problem for naturalness. The gluino, in particular, is critical for naturalness for two reasons: first, it is the most abundantly produced, due to its large color charge. Second, the gluino, if heavy, pulls upward the masses of all the colored sparticles including the stop, which in turn pulls up the Higgs and the weak scale. This behavior is illustrated in Fig.~\ref{fig:gluino}, where we see that even with a decade of RG running {\it a heavy gluino sucks} the squark masses up to within a factor of two of the gluino mass and the up-Higgs soft mass to within a factor of seven. 

\begin{figure}[t!] 
 \begin{center}
 \includegraphics[height=7cm]{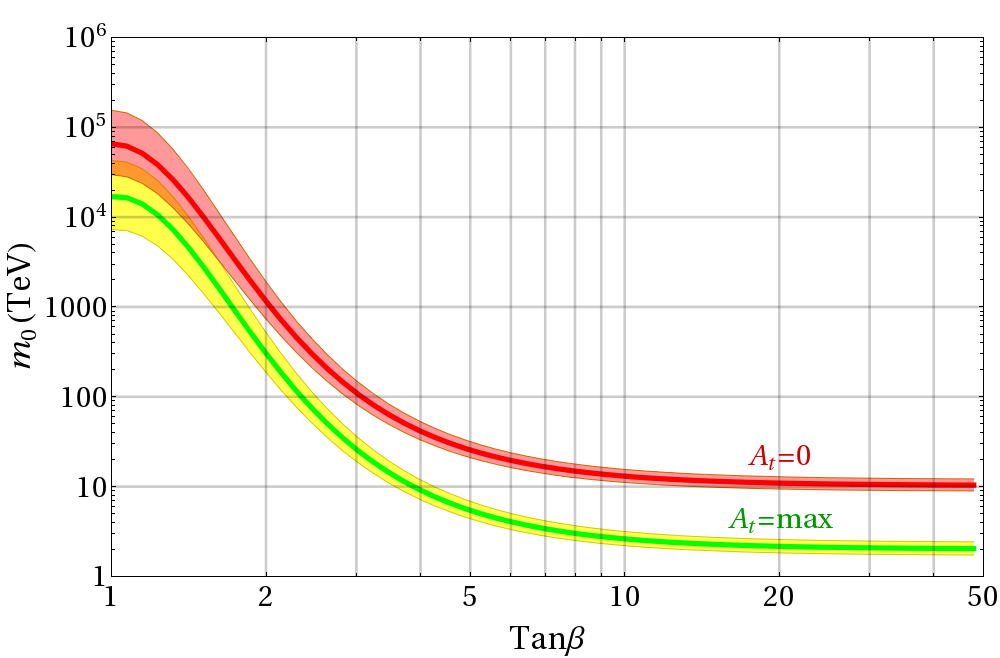}
 \caption{The scalar mass scale in Split Supersymmetry as a function of $\tan \beta$ for a Higgs mass fixed at 125.5~GeV for no and maximal stop mixing. The 1$\sigma$ error bands coming from the top mass measurement (which dominate over other uncertainties) are also shown.}
 \label{fig:tanbetavsms}
 \end{center}
\end{figure}

\begin{figure}[t!] 
 \begin{center}
 \includegraphics[width=0.6\textwidth]{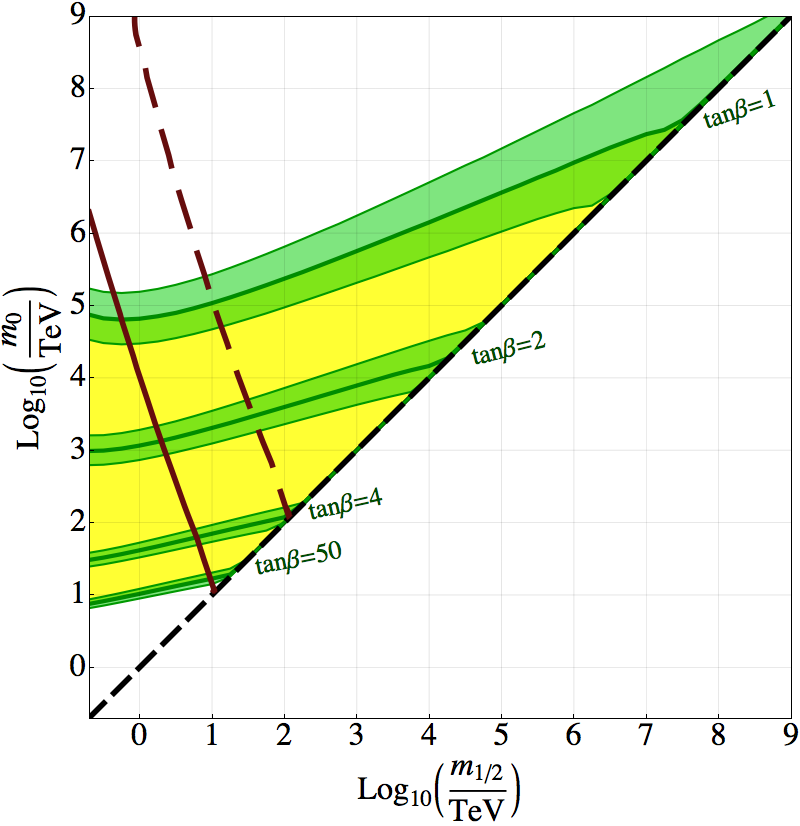}
 \caption{The Higgs mass (here chosen to be 125.5~GeV) constrains the scalar and fermion masses to be in the shaded region, for varying $\tan \beta$. The green bands are the 1$\sigma$ error from the top mass measurement for the given value of $\tan \beta$. Gauge coupling unification constrains the parameters to be to the left of the solid bordeaux (1$\sigma$) or dashed bordeaux (2$\sigma$) lines as described in the text. This plot was generated using the results of \cite{strumia}.  }
  \label{fig:m0mu}
 \end{center}
\end{figure}

In Fig.~\ref{fig:tuning} we show the amount of fine-tuning in Natural SUSY theories where only higgsinos, stops and gluino are light \cite{naturalsusy}. We consider only the contributions coming from the higgsino, the stop and the gluino and assume the higgs mass of 126~GeV is generated at tree-level. This gives  a lower bound to the amount of tuning, which applies to any SUSY model, including MSSM, NMSSM, $\lambda$SUSY and models with non-decoupling D-terms. We plot the product of two tunings: one is the usual tuning for the electroweak vev and the other is the tuning required to keep $\tilde t_1$ light when  $m_{\lambda_3}\gg m_{\tilde t_1}$.
From Fig.~\ref{fig:tuning} we see that, given the current gluino search bounds \cite{naturalsusybounds}, there is at best $\sim10\%$ fine-tuning in the theory even when there is a mere order of magnitude between the messenger scale $\Lambda$ where the soft terms are generated, and the sparticle mass. The minimal tuning allowed in the theory deteriorates as $\Lambda$ is increased. It is less than $10\%$ already with a loop hierarchy between the sparticles and $\Lambda$ (as in low-scale gauge mediation) and drops below $1\%$ in high scale SUSY breaking models (as in gravity or anomaly mediation). The gluino RG effects become stronger as $\Lambda$ is pushed up and it gets harder to have a stop much lighter than the gluino.

 The bounds on the tuning from current direct stop searches are not competitive with the gluino ones, and thus do not pose a significant constraint on the parameter space. When $m_{\lambda_3}\gg m_{\tilde t_1}$, additional tuning is required because of the large correction to the stop mass from the gluino. Making the LSP heavier than 400~GeV to evade the gluino bounds does not improve the situation; a heavy LSP implies a large $\mu$-term which increases the tree-level tuning of the theory. Fig.~\ref{fig:tuning} 
finally shows that the small window left for naturalness in SUSY will be probed already by the end of the 8~TeV LHC run, when the gluino searches are pushed above 1.5-1.8~TeV mass range.

The absence of evidence for sparticles suggests that either low-energy SUSY theories have to be tuned, or sparticles are absent from the weak scale altogether. Why, then, does supersymmetric unification work so well if the sparticles responsible for it are not present? An answer to this question comes from Split SUSY \cite{ad04,gr04}, a theory motivated by the multiverse. In Split SUSY, scalar sparticles are heavy---at the SUSY breaking scale $m_0$---whereas fermions (gauginos and higgsinos) are lighter as they are further protected by the R-symmetry whose breaking scale can be lower than $m_0$. Choosing the fermion masses near a TeV, as dictated by the WIMP  ``miracle", reproduces successful unification independent of the masses of scalar sparticles. So in Split only the gauginos and higgsinos may be accessible to the LHC, whereas the scalar masses can be anywhere between the GUT and the weak scale.


This uncertainty in $m_0$, which has been blurring the phenomenology and model building of Split, has come to an end with the discovery of the Higgs~\cite{higgsdiscover}. The Higgs mass $m_h$ correlates with $m_0$ \cite{ad04,gr04} as shown in Fig.~\ref{fig:strumia}  \cite{strumia}, and for $m_h=125.5$ GeV the scalar sparticle masses are in the range from $10$ TeV to $10^5$ TeV. Figure \ref{fig:tanbetavsms} exhibits the relation between $m_0$ and $\tan \beta$ fixing the Higgs mass to its observed value. Note that heavy scalar masses above $10^3$ TeV are only possible for a limited range of small $\tan \beta \lesssim 2$, whereas any value of $\tan \beta \gtrsim 3$ implies scalar masses less than 100 TeV. This is a potentially exciting low mass range suggesting that the gauginos and higgsinos may be LHC-accessible, independently of the  WIMP miracle. The reason is that in many models of SUSY breaking the gauginos are much lighter than the scalars, as they are protected by $R$-symmetry. In fact one has to work hard to ensure that the SUSY and $R$-breaking scales coincide. In simple models of anomaly mediation, for example, the gauginos are one loop lighter than the scalars. Indeed, the range of $m_0$ indicated by the Higgs mass is suggestive of a one- or two-loop separation between scalars and gauginos.

Another constraint comes from unification, which prefers low values for the $\mu$ parameter. This is underlined in Fig.~\ref{fig:m0mu}, where we show the correlation between the scalar and the fermion masses $m_{1/2}$, assuming that the higgsino mass $\mu$ and the gaugino masses $m_{\lambda_{1,2,3}}$ are equal\footnote{The higgs mass prediction is still valid for $m_{1/2}={\rm Max}(\mu, m_{\lambda_i})$ as the one-loop RG contributions to the Higgs quartic from the fermions require both the higgsinos and the gauginos. The limits from unification are also approximately valid for $m_{1/2} = \mu > m_{\lambda_i}$ since the higgsinos are the most important degrees of freedom for unification.}. The observed value of the Higgs mass chooses the shaded region, whereas unification prefers us to be to the left of the solid ($1 \sigma$) or the dashed ($2 \sigma$) bordeaux line. Here $\sigma$ is chosen to be the magnitude of the two-loop corrections, which is a reasonable model-independent proxy for the one-loop threshold corrections. Unification prefers values of $\mu$ less than 100 TeV, again raising the hope for LHC observability.

The RG running for the scalar soft terms in Split supersymmetry has received almost no attention in the literature (with the notable exception of \cite{Ibarra}); it was deemed uninteresting because the scalars, other than the Higgs, are not observable. In the next section we study the scalar soft terms and show that they are crucial for consistent model building.  The scalar masses, especially those of the stop and Higgs, tend to go tachyonic due to the smallness of the gaugino masses. This has implications for electroweak symmetry breaking and puts significant constraints on the spectrum that we discuss in detail in the next section. In section 3 we discuss a number of mini-Split models and in section 4 we classify the various mini-Split models according to the phenomenological signatures.



\section{A Guide to Tuning the Electro-Weak Scale in Split SUSY}
\label{EWSBsection}

\begin{figure}[t!]
  \centering
  \hspace{-0.4cm}
  \begin{minipage}[c]{0.35\textwidth}
    \centering
    \begin{tabular}{||c|c||}
    \hline  \hline
     & $\mu\ll m_{H_d}$\\ 
    $\textcolor{blue}{m_{H_u}^2>0}$ &$\tan\beta\simeq m_{H_d}/m_{H_u}$\\ \cline{2-2}
    & $\mu \gtrsim m_{H_d}$\\ 
     &$\tan\beta\simeq 1$\\
     \hline 
     & $\mu\sim |m_{H_u}|$\\ 
    $\textcolor{red}{m_{H_u}^2<0}$ &$\tan\beta>$ few \\ \cline{2-2}
    & $\mu > m_{H_d}$\\ 
     &$\tan\beta\simeq 1$\\
     \hline  \hline
    \end{tabular}
  \end{minipage}
  \hspace{1cm}
  \begin{minipage}[c]{0.55\textwidth}
\includegraphics[width=1.04\textwidth]{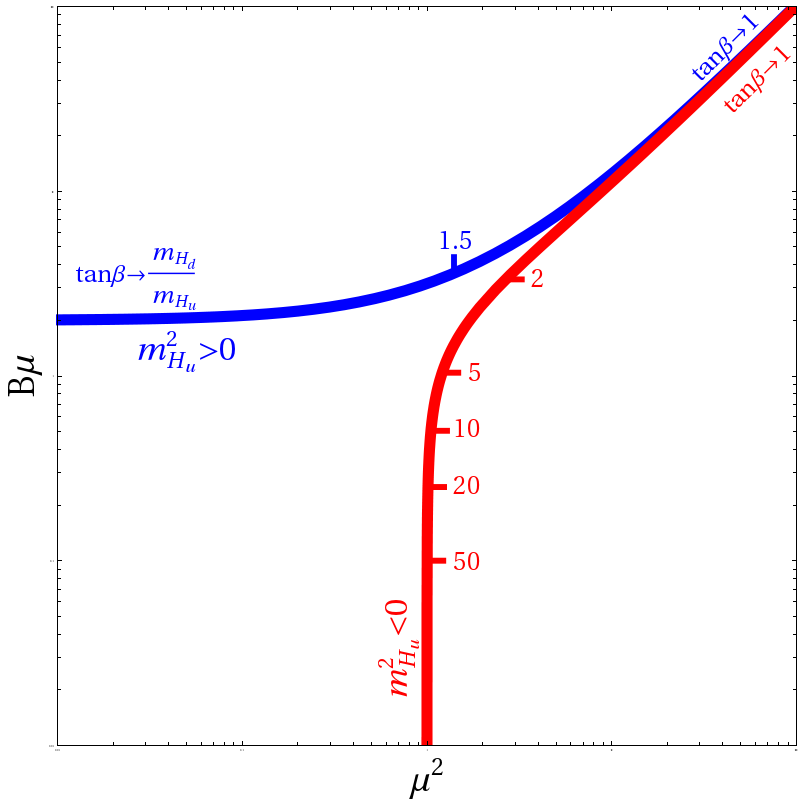}
  \end{minipage}
  \caption{The constraints on the Higgs and higgsino parameters from the tuning of the EW scale are summarized in the table (left) and in the plot (right). In the plot, the tuning of the EW scale constrains $\mu$ and $B_\mu$ to
  lie on the blue (red) curve if $m_{H_u}^2>0$ ($m_{H_u}^2<0$). }
  \label{fig:EWSB}
\end{figure}

In concrete models of Split supersymmetry, successful tuning of electroweak symmetry breaking (EWSB) is not automatic. It requires that the Higgs sector parameters at the scalar mass scale satisfy the following relation:
\be
\label{EWSBcondition}
\det \left( \begin{array}{ccc}
|\mu|^2 + m^2_{H_u} & -B_{\mu} \\
 -B_\mu^*  & |\mu|^2 + m^2_{H_d}  
 \end{array} \right) \approx 0 \,, \qquad 
 \tan \beta = \sqrt{\frac{m_{H_d}^2+|\mu|^2}{m_{H_u}^2+|\mu|^2}}\,,
\ee 
where both the diagonal elements of the mass matrix must be positive to avoid unacceptably large values for $\tan\beta$. Moreover the condition $\tan\beta\geq 1$ implies $m_{H_d}^2\geq m_{H_u}^2$.

We can thus distinguish two different cases: $m^2_{H_{u}}$ positive or negative. In the first case, the higgsinos can be well below the scalar mass scale so that
\be
 |B_{\mu}|^{2} \approx m_{H_{u}}^2m_{H_{d}}^2 \hspace{1cm} \text{and} \hspace{1cm}
 \tan \beta\approx \frac{m_{H_d}}{m_{H_u}}.
\ee
This pattern of EWSB can allow for models where higgsinos and gauginos are at the same scale, as was suggested in the original Split papers. 

When $m^2_{H_{u}} < 0$, on the other hand,  the higgsinos must be at least as heavy as the scalars, aspects of which were considered in \cite{Cheung:2005ba}. EWSB can be achieved through either tuning the $\mu$ term against $m^2_{H_u}$ with small $B_{\mu}$ and large $\tan \beta$, or by requiring that $|\mu|^2 \approx |B_{\mu}| \gtrsim |m^2_{H_u}|$ with $\tan \beta \sim 1$. The latter case is disfavored by unification due to the largeness of $\mu$, as shown in Fig.~\ref{fig:m0mu}.
Both of these cases represent a clear departure from ordinary Split phenomenomenology. When the higgsinos are much heavier than the gauginos, the mixing in the electroweakino sector is suppressed unless the bino and wino masses are tuned to be degenerate. Absent this tuning, well-tempered dark matter \cite{welltempered} is no longer a possibility and the only option for a calculable DM candidate is a wino LSP whose mass is constrained to be around $\sim 2.5$ TeV. In the presence of another DM particle in the theory, like the QCD axion, the wino can be much lighter and accessible at the LHC. Another consequence of decoupling the higgsinos from the spectrum is the near-degeneracy of the charged and neutral wino states, which are now separated only by 155-175 MeV in mass due to electroweak corrections \cite{Feng:1999fu}. As we will discuss further in section 4, chargino decays to the neutralino in this case produce a displaced pion and are subject to dedicated LHC searches \cite{ATLAS:2012ab} motivated by anomaly-mediated scenarios. 

If the bino is the LSP, it is always overproduced~\cite{Dimopoulos:1990kc}~and its abundance must be diluted via, e.g., decay to a much lighter gravitino or axino. For a 100 GeV bino to decay before BBN, the scale of SUSY breaking has to be $\sqrt{F}\lesssim 10^8$~GeV or the axion decay constant $f_a \lesssim 10^{11}$~GeV.

When the $\mu$ term is large, there is also an irreducible contribution to the wino and bino masses~ \cite{ghergiudwells} 
\begin{equation}
\label{eq:mucorr}
\delta M_1=\frac{3}{5} \frac{\alpha_1}{4 \pi} \mu \frac{2 B_{\mu}}{m_A^2-|\mu|^2}\log \left( \frac{m_A^2}{\mu^2} \right) \quad {\rm and}\quad
\delta M_2= \frac{\alpha_2}{4 \pi} \mu \frac{2 B_{\mu}}{m_A^2-|\mu|^2}\log \left( \frac{m_A^2}{\mu^2} \right)
\end{equation}
for the bino and the wino respectively, which suggests that the electroweakinos can be separated from the higgsinos by at most a factor of $\sim 4 \pi \tan \beta \, /\alpha_i $. In this case the wino and bino can be lifted, resulting in a gluino LSP. This possibility is disfavored by unification and is experimentally excluded unless the gluino can decay into an axino or gravitino before BBN.

The constraints on the Higgs and higgsino parameters from the tuning of the EW scale are summarized in Fig.~\ref{fig:EWSB}. It is clear from the discussion above that even in Split model building, just like in low energy SUSY, special attention has to be given to the way $\mu$ and $B_{\mu}$ are generated. A solution to the $\mu$ and $B_{\mu}$ problem is in order when  $m^2_{H_{u}} < 0$. On the other hand, a light higgsino requires $m_{H_u}^2 > 0$, which is nontrivial in Split SUSY after taking into account the RG running of the scalar masses, as we will discuss further in the next section.

\subsection{Tachyonic scalar masses from RG running}
\label{scalarRGEsection}
\begin{figure}[t] 
 \begin{center}
 \includegraphics[width=0.48\textwidth]{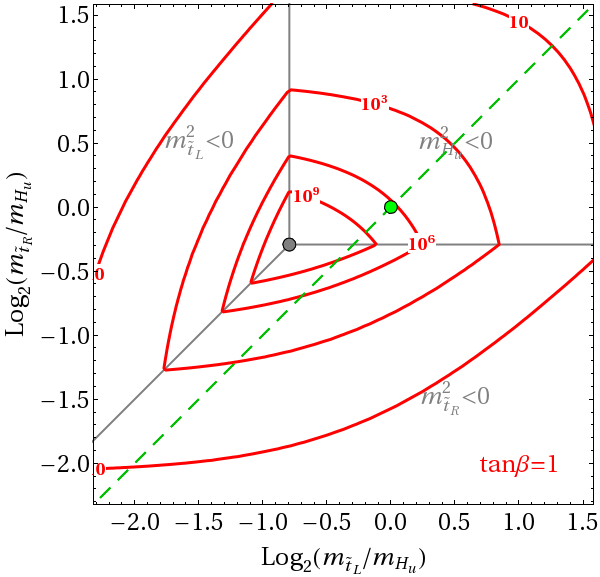}
  \includegraphics[width=0.48\textwidth]{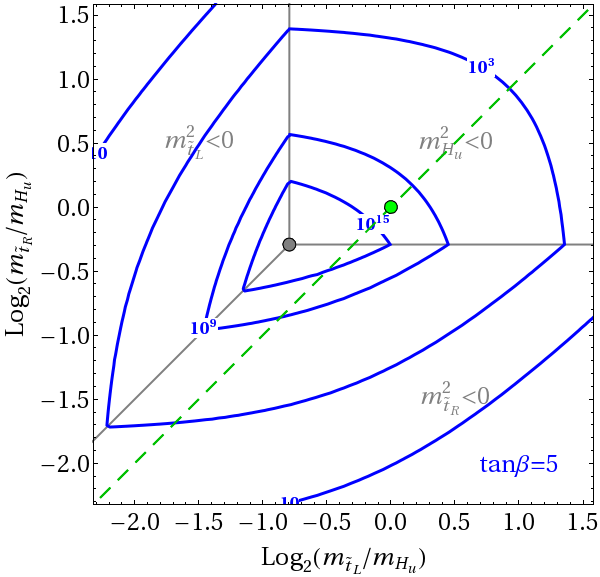}
 \caption{Contour plots of the amount of running required for either of $m_{\tilde t_L}^2, m_{\tilde t_R}^2, \text{or } m_{H_u}^2$ to become negative in the $\frac{m_{\tilde t_L}}{m_{H_u}}$ vs $\frac{m_{\tilde t_R}}{m_{H_u}}$ plane for $\tan \beta= 1$ (left) and $\tan \beta =5$ (right). The running of the top Yukawa is not included. The UV fixed point is also shown (light grey) as well as the point which corresponds to universal boundary conditions (light green). The dashed green line shows when $m_{\tilde t_L}^2 = m_{\tilde t_R}^2$.}
 \label{tachyons}
 \end{center}
\end{figure}

Taking into account RG effects reveals another important feature of Split SUSY: the scalar masses of $\tilde t_{L(R)}$ and $H_u$ can easily run negative. The main reason is the absence of the gaugino contribution in the running that normally protects these masses in the MSSM. In addition, $\tilde t_{L(R)}$ and $H_u$ receive large radiative corrections coming from the top yukawa:
\be
\frac{d m_i^2 }{dt} = c_i X_t  \equiv c_i \frac{ |y_t|^2}{8 \pi^2} (m^2_{H_u}+m^2_{{\tilde t_L}}+m^2_{{\tilde t _R}})\,,
\ee  
where $c_i$ takes the values 1, 2 or 3 for $\tilde t_{L}$, $\tilde t_{R}$ and $H_u$, respectively. If there is a large hierarchy between these three masses at the messenger scale, then radiative corrections can push the smallest of the masses negative even with  minimal   running. This is because $X_t=0$ is the IR fixed point of the RGEs. The amount of running required for any one of the masses to become tachyonic as a function of the ratio between the masses is shown in Fig.~\ref{tachyons} for two different values of $\tan \beta$.  The dependence on $\tan \beta$ enters through the change in the top Yukawa coupling. There is a large part of the parameter space that is excluded because the colored scalars become tachyonic. In addition, if universal boundary conditions are imposed for the scalars at the GUT or Planck scale, $H_u$ will very quickly run negative. In this case the higgsinos are now required to be at the scalar mass scale to have successful EWSB in these models, resulting in tension with gauge coupling unification as discussed earlier.

There is one more fixed point of the RGEs, which is in the UV. This can be easily seen by looking at the evolution of the mass ratios (ignoring the running for the top yukawa):
\be
\frac{d}{dt} \left(\frac{m^2_{H_u}}{m^2_{{\tilde{t}_L}}}\right) \approx \left(3-\frac{m^2_{H_u}}{m^2_{{\tilde{t}_L}}}\right) \frac{X_t}{m^2_{{\tilde{t}_L}}}, \hspace{2cm} 
\frac{d}{dt} \left( \frac{m^2_{{\tilde{t}_R}}}{m^2_{{\tilde{t}_L}}}\right) \approx \left(2-\frac{m^2_{{\tilde{t}_R}}}{m^2_{{\tilde{t}_L}}}\right) \frac{X_t}{m^2_{{\tilde{t}_L}}}\,.
\ee
These equations have fixed points for $m^2_{H_u} =3 m^2_{{\tilde{t}_L}}$ and $m^2_{{\tilde{t}_R}}=2{m^2_{{\tilde{t}_L}}}$, respectively. Thus the RGEs push the square of the three masses of $\tilde t_{L}$, $\tilde t_{R}$ and $H_u$ to the UV fixed ratio $1:2:3$, respectively (see Fig.~\ref{tachyons}). The closer the UV boundary conditions are to these ratios, the slower the running is to the IR fixed point of $X_t=0$. In order for the masses to be all positive at low scales, the initial conditions have to be close to the UV fixed point.

In general, there is another contribution in the running that can destabilize more scalars coming from the hypercharge D-terms:
\be
\frac{d m^2_i }{dt} \supset \frac{6}{5} Y_i \frac{\alpha_{{Y}}}{4 \pi} {\rm{Tr}}(Y m^2)
\ee
This contribution can easily be zero if the scalar soft masses satisfy $m_{H_u}^2=m_{H_d}^2$ and GUT relations. Of course, RG effects will generate a non-zero contribution from this term that is equivalent to 2-loop effects and thus suppressed.  If $m_{H_u}^2 \neq m_{H_d}^2$ or the GUT relations are not satisfied, in general the amount of running required for scalars to become tachyonic will be reduced. In this case the full effects of RG running should be included in all Split models given the large number of soft parameters involved. 

Taken together, the requirements of electroweak symmetry breaking and the effects of RG evolution have considerable implications for the low-energy phenomenology of Split supersymmetry. Even though the scalars are inaccessibly massive, they may leave an imprint on the spectrum of light neutralinos through the fine-tuning of the weak scale. Ultimately, the precise interplay between scalar soft masses, RG effects, EW fine-tuning, and LHC phenomenology depends on the details of the UV model.

\section{Models of Mini-Split}

\subsection{Anomaly Mediation}

Perhaps the simplest scenario for a mini-Split spectrum is un-sequestered anomaly-mediated supersymmetry breaking (AMSB) \cite{amsb}. In this case scalar masses are dominated by gravity-mediated contributions at $\mathcal{O}(m_{3/2})$, while gaugino masses are protected by an $R$-symmetry and arise at one loop via anomaly mediation,
\begin{equation}
m_{\lambda_i} = \frac{\beta(g_i)}{2 g_i} m_{3/2}\sim \frac{b_i g_i^2}{16 \pi^2} m_{3/2} \, ,
\end{equation}
where $b_i$ is the beta function coefficient for the relevant gauge group. This naturally leads to a loop-order splitting between the scalars and gauginos. The values of $\mu$ and $B_\mu$, on the other hand, depend on the details of the hidden sector and its interactions with the Higgs multiplets. There are two natural cases: 
\begin{itemize}
\item 
$B_\mu\sim |\mu|^2\sim m_{3/2}^2$: where the tuning of EWSB can be achieved through either $\mu$ or $B_\mu$ and $m_{H_u}^2$ can run negative. This can be achieved either by fixing $\mu\approx m_{3/2}$, which generates $B_\mu\sim \mu m_{3/2}$, or by the Giudice-Masiero mechanism triggered by the operators $XX^\dagger H_u H_d/M_{Pl}^2$ and $X^\dagger H_u H_d/M_{Pl}$ with $X=M_{Pl}+F \theta^2$.
\item
$B_\mu \sim m_{3/2}^2 > \mu^2$: where the tuning of EWSB must be imposed through $B_\mu$, $m_{H_u}^2$ must not run negative and the higgsino can be light. This case can be realized again with the Giudice-Masiero mechanism giving an $R$-charge to $X$ different from 2 and a vev to $X$ smaller than $M_{Pl}$.
\end{itemize}

Given that the scalar spectrum is gravity mediated, without an alignment mechanism,
$O(1)$ flavor violations are expected and FCNCs bound the scalar masses to lie above few$\cdot10^3$~TeV~ \cite{Gabbiani:1996hi}.
Despite the relative heaviness of the scalars, mini-Split AMSB suffers from the persistence of flavor problems.

A viable AMSB model is subject to a variety of additional constraints beyond the flavor problem. The effects of RG evolution on the soft spectrum are quite important given the high scale of mediation. While $SO(10)$ gauge invariance of the soft spectrum may be invoked to guarantee that contributions to RG flow proportional to the hypercharge trace vanish, the contributions from the Yukawa may still lead to unwanted tachyonic states. 

Combining all the information above, we can envision two different scenarios:
\begin{enumerate}
\item \emph{Heavy AMSB Mini-Split:} The scalars are around $10^4$~TeV, which automatically ensures a solution to the flavor problem. The gauginos are lighter by only one loop, around $10^2$~TeV, and are out of the LHC reach. Nevertheless, gauge coupling unification favors a $\mu$ term that is much lighter than the scalars and the higgsinos may be produced at the LHC. In addition, EWSB must be tuned via the $B_\mu$ term, $\tan \beta \sim 1$, and particular boundary conditions have to be imposed so that $m_{H_u}^2>0$.

\item \emph{Light AMSB Mini-Split:} In this case the scalars are around $10^2$~TeV and extra model-building is required to address the flavor problem. The $\mu$ term can be naturally at the same scale in agreement with gauge coupling unification. Its contribution to the tuning of EWSB relaxes the requirement on the sign of $m_{H_u}^2$. Finally, the gauginos are at the TeV scale and may be within the LHC reach. Their spectrum may deviate from the pure anomaly mediated one due to the radiative corrections from the $\mu$ term in Eq.~(\ref{eq:mucorr}).
\end{enumerate}

The LHC prospects thus depend on the origin of $\mu/B_\mu$ and the solution to the flavor problem. If there is no mechanism for alignment of the soft parameters, $m_{3/2}$ must be so large that the gauginos are beyond the reach of the LHC. The higgsinos may remain light, in which case the LHC phenomenology is that of the minimal model, discussed in section \ref{sec:pheno}. When the flavor problem of anomaly mediation is addressed, gauginos can be LHC accessible with the gluino production and decay providing a measurement of the scalar masses.

\subsection{$U(1)'$ Mediation}

The persistent flavor problem of anomaly mediation motivates  flavor-blind models for mini-Split, such as  those in which SUSY breaking is communicated via gauge interactions. The mediating interactions may be those of the MSSM or of an additional gauge group. Here we consider a model in which SUSY-breaking is mediated by an extra abelian gauge multiplet, which naturally gives a mini-Split spectrum and exemplifies the constraints discussed in sections \ref{intro} \& \ref{EWSBsection}.

In particular, we extend the SM gauge group by an additional $U(1)'$ under which the vector-like messengers, $S$ and $\bar{S}$, are charged. For simplicity, we take $U(1)'$ to be a linear combination of $U(1)_{B-L}$ and $U(1)_{Y}$ which we parametrize by the angle $\theta$:
\be
U(1)'=\cos(\theta)\, U(1)_{B-L}+ \sin(\theta)\, U(1)_{Y}\,.
\ee
This choice is well-motivated by anomaly cancellation as well as by Grand Unification. The moment the unifying group is extended beyond $SU(5)$ there are easily additional $U(1)$ groups that may be broken above the weak scale. For definiteness we choose the messengers' charge to be the same as the one of the right handed neutrinos. As in gauge mediation,  the scalar masses arise at two loops:
\be
m_{0}^2=\left(\frac{\alpha'}{4 \pi}\right)^2 q'^2_{0} q'^2_{{S}} \left(\frac{F}{M_{{S}}}\right)^2\,, 
\ee
where $\alpha'=\frac{g'^2}{4 \pi}$, $q'_{0}$, and $q'_{{S}}$ are the $U(1)'$ coupling strength, the MSSM scalar and messenger charge under the $U(1)'$. $F$ is the F-term SUSY breaking vev and  $M_{{S}}$ is the supersymmetric messenger mass. There can be an additional contribution to the scalar masses coming from D-term SUSY breaking,
which can be of the same order or smaller than the F-term contribution:
\be
\delta m^2_{0}= g'^2 q'_{0} \xi_{FI}\,.
\ee
\label{U(1)Split}
\begin{figure}[t!] 
 \begin{center}
 \includegraphics[width=0.7\textwidth]{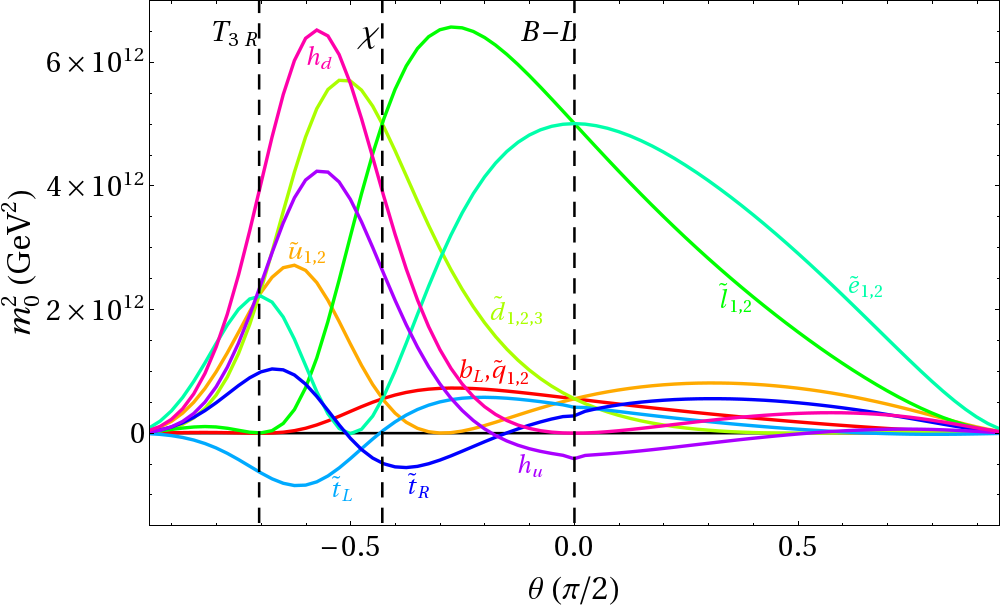}
  \includegraphics[width=0.7\textwidth]{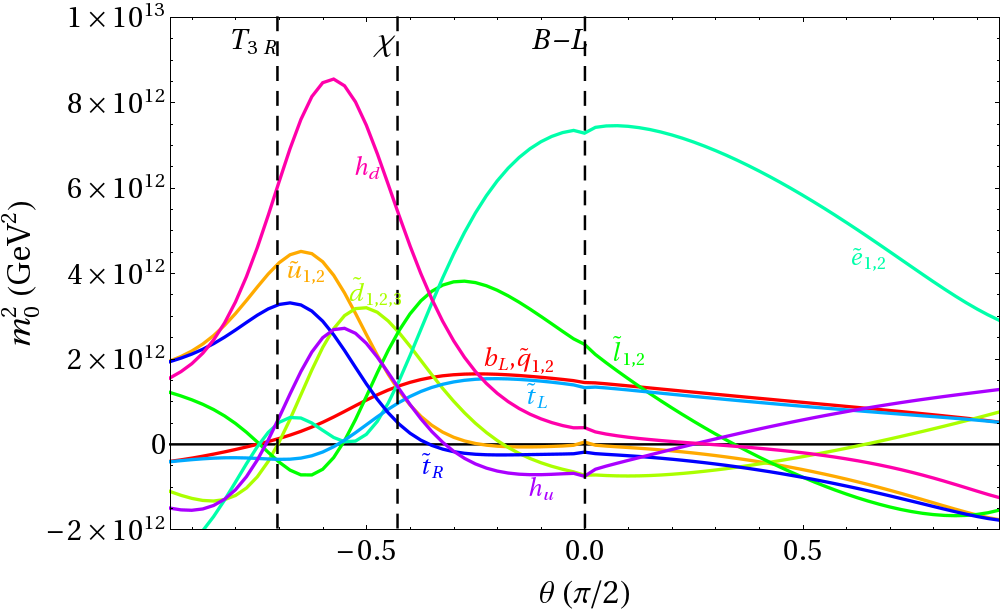}
 \caption{Scalar spectrum with zero (top) and non-zero (${g'}^2\xi_{FI}=5 \cdot 10^6$~TeV$^2$) D-terms (bottom) as a function of the mixing angle $\theta$ between $U(1)_{Y}$ and $U(1)_{B-L}$. We fixed $M_S=3\cdot10^{8}$~TeV, $\sqrt{F}=3\cdot10^{7}$~TeV, $\alpha'(M_S)^{-1}=40$,  $\alpha_{\lambda_{u,d}}=100\, {q}_{H}^{\prime2} q_S^{\prime2} (\alpha'/4\pi)^2$. With vertical dashed lines we denote the value of $\theta$ which gives rise to the left-right ($U(1)_{ T_{3R}}$), high-scale ($U(1)_\chi$) and low-scale ($U(1)_{B-L}$) $B-L$ models.}
 \label{fig:scalarswithDterms}
 \end{center}
\end{figure}

For the MSSM gauginos the situation is very different from ordinary gauge mediation. Since the messengers carry no SM charge, gauginos feel R- and SUSY-breaking at the three-loop level \cite{Craig:2012di}:
\be
m_{\lambda_i}=\frac{1}{8 \pi^3} \alpha_i \alpha'^2 q'^2_{{S}} \sum_Q C_i(Q) q'^2_{Q}  \frac{F}{M_{{S}}} \,,
\ee
where $i={1,2,3}$ enumerates the SM gauge groups, $C_i(Q)$ is the gauge group Casimir, and the sum runs over all SM fields. 
As a result, there is a two-loop hierarchy between the gauginos and the scalars:
\be
\frac{m_{\lambda_i}}{m_{ 0}}=\frac{\alpha' \alpha_i}{2 \pi^2}\frac{q'_{{S}}}{q'_{0}} \sum_Q C_i(Q) q'^2_{Q}\,.
\ee
For an $\alpha'$ similar to SM couplings and given that $\sum_Q C_i(Q) q'^2_{Q}\sim 50$, we find that, when the scalars are at $10^3-10^4$~TeV, the gauginos are at the TeV scale. Note that in the above discussion $|F|^{1/2}$ has to be smaller than  $\sim10^{11}$~GeV in order for the anomaly-mediated contribution to the gaugino masses to be subdominant and larger than $\sim10^{10}$~GeV in order for our expansion in $\frac{|F|}{M^2_{{S}}}$ to be perturbative.

\subsubsection{Scalar Mass Spectrum and Stability}

In Fig.~\ref{fig:scalarswithDterms} we present the scalar spectrum including the RG effects from $M_{{S}}=3\cdot 10^{8}$~TeV down to $m_{0}$ without and with including a small D-term contribution, respectively, as a function of the angle $\theta$. We have assumed that $U(1)'$ is broken right below $M_{{S}}$ so we have not taken into account the corresponding running effects. Both figures illustrate how easily the scalar masses of $\tilde t_{L(R)}$ and $H_u$ can run negative, as discussed in section \ref{scalarRGEsection}. Without any contribution from the D-terms, one of these scalars becomes tachyonic for any value of $\theta$. When adding a D-term contribution
of the same order of the scalar masses, we find that there is a small region around $U(1)_\chi$, the high-scale $B-L$, where all scalars have positive mass squared. This is because the SM particle charges under $U(1)'$ are such that the D-terms push the UV values of the $\tilde t_{L(R)}$ and $H_u$ squared masses closer to the UV fixed ratio of $1:2:3$ and they do not run enough to become negative. 
The $U(1)_\chi$ model is also attractive from the point of view of unification and an attractor for the RGE evolution of the mixing angle $\theta$. The other region that is allowed with the addition of (different) opposite sign D-terms and it is not shown in Fig.  \ref{fig:scalarswithDterms}, is around $\theta \sim \frac{3 \pi}{8} $ but it requires that $U(1)'$ and $U(1)_{Y}$ are almost parallel at the high scale. 
The region around (low-scale) $U(1)_{B-L}$, where $m^2_{H_u}$ runs tachyonic without the addition of D-terms (as can be seen in Fig. \ref{fig:scalarswithDterms}) is also phenomenologically interesting as we will discuss in the next section.

\subsubsection{EWSB}
\label{EWSBU(1)}

The size of the $\mu$-term in this model is closely related to how electroweak symmetry breaking (EWSB) is tuned. In addition, we have to take into account the requirement on $\tan \beta$. As previously discussed, the two loop hierarchy between the scalars and the gauginos pushes the scalars at least as heavy as $10^3$~TeV, unless $\alpha'$ runs strong well before the GUT scale.  We are thus forced to consider models where $\tan \beta \sim 1$, as can be seen from Fig.~\ref{fig:tanbetavsms}.  
The correct EWSB pattern can thus be achieved in this model in two different ways (see Fig.~\ref{fig:EWSB}):
\begin{itemize}
\item when $|B_\mu| \simeq m^2_{H_{u}} \simeq m^2_{H_{d}} > |\mu|^2$, or
\item when $|B_\mu| \simeq |\mu|^2  \gtrsim |m^2_{H_{u(d)}}|$
\end{itemize}

In the first case, we can turn the usual $\mu-B_\mu$ problem  of gauge mediation to an advantage. By adding a pair of $\mathbf {5 \oplus \bar 5}$ messenger fields which do not couple to the SUSY breaking spurion, we can write direct couplings of the Higgs fields to $S$ and $\bar S$:
\be 
W\supset \lambda_{u} H_{u} D S + \lambda_d H_{d} \bar D \bar S\,.
\ee
These couplings generate both a $B_\mu$ and a $\mu$ term at one loop \cite{lopsided}:
\be
B_\mu \varpropto \frac{\lambda_u \lambda_d}{16 \pi^2}  \left | \frac{F}{M_{{S}}} \right| ^2 \,,
\ee
\be
\mu \varpropto \frac{\lambda_u \lambda_d}{16 \pi^2}  \left | \frac{F}{M_{{S}}} \right|\,.
\ee
The ratio between $\mu$ and $B_\mu$ is thus:
\be
\frac{|\mu|^2}{B_\mu} \varpropto \frac{\lambda_u \lambda_d}{16 \pi^2}\,.
\ee

From Eq. \ref{EWSBcondition} we can infer the size of the $\mu$ term. We assume that the contribution to the soft Higgs masses coming from the direct coupling to the messengers is subdominant and we find that EWSB is achieved when
\be
\frac{\lambda_u \lambda_d}{16 \pi^2} \sim \left( \frac{\alpha'}{4 \pi}\right)^2 \,.
\ee


This suggests that the $\mu$-term is also much below the scalar masses and it is very close to the TeV scale. There is no two-loop contribution to the gaugino masses coming from the direct coupling of the Higgs to the messengers due to gaugino screening~\cite{aglr, cohen}. Taking into account the threshold corrections from the $\mu$ term to the gaugino mass we find that the bino is the LSP. In particular, for $U(1)_{\chi}$ mediated SUSY breaking, 
the ratio between the fermion masses is approximately $m_{\lambda_1}:m_{\lambda_2}:m_{\lambda_3}:\mu \approx 1:3:12:13$, 
and the gluino can easily be within the reach of the LHC. 

Dark Matter in this model cannot be the bino---it is always overproduced by at least a factor of 50, as the main annihilation channel is $t$-channel higgsino exchange which is p-wave and the cross-section is suppressed by $\mu^{-2}$~\cite{welltempered}. Decays to the gravitino are not enough to dilute the bino abundance, since the gravitino, which is the true LSP, is between 10 and 100 GeV so an additional dilution mechanism is required. This can be provided by possible decays to the light QCD axino, which due to BBN constraints has to have an axion decay constant smaller than $10^{11}$~GeV.

When $m^2_{H_u} < 0$ as is the case for pure $B-L$ mediation, we need to generate a $\mu$ term that is of order $\sqrt{|B_\mu|}$. This means that the higgsinos are now at $\sim10^3$~TeV, in tension with unification as can be seen from Fig.~\ref{fig:m0mu}. Nevertheless it is still a viable and phenomenologically interesting possibility. An example of how $\mu \sim \sqrt{|B_{\mu}|}$ can be achieved in gauge mediation is given in \cite{GKR}. With two sets of messenger fields, $S_1/\bar S_1$ and $S_2/\bar S_2$ and an overall singlet $N$, we can write the following technically-natural superpotential:
\be
W \supset  \lambda N H_u H_d + \frac{1}{2} M_N N^2 + (m + \xi N) \bar S_1 S_2 + X (\bar S_1 S_1 + \bar S_2 S_2)\,,
\ee
where $X = M_{ S} + F \theta^2$ is the spurion that breaks SUSY. There is a $\mu$ term generated at one loop:
\be
\mu =  \frac{\lambda \xi }{16 \pi^2} \frac{m^\dag}{M_N} \frac{F^\dag}{ M_S^\dag}.
\ee
Contributions to the $B_\mu$ term are generated at both one and two loops; the one-loop contribution is proportional to $m^2 / M_S^2$ and may be rendered negligible for $m \ll M_S$, while the two-loop contribution is 
\be
B_\mu = - \frac{2 \lambda \xi^3}{(16 \pi^2)^2} \frac{m^\dag}{M_N} \frac{|F|^2}{|M_S|^2}.
\ee

Then we have for the tuning condition, using field redefinitions to make $B_\mu$ positive, and assuming that $|m^2_{H_u}| \ll |\mu|^2$:
\be
B_\mu / |\mu|^2 = 2 \frac{ \xi }{ \lambda} \frac{M_N^\dag }{m}\approx1\,.
\ee
As already discussed in Sec. \ref{EWSBsection}, because both $\mu$ and $B_\mu$ are large we have to take into account their contribution to the bino and wino masses. Doing so, we find that they are heavier than 10~TeV with the gluino being the LSP, which cannot be stable and has to decay before BBN to e.g. the axino or gravitino.

\subsubsection{Hybrid Mediation}
\begin{figure}[t!] 
 \begin{center}
 \includegraphics[width=0.65\textwidth]{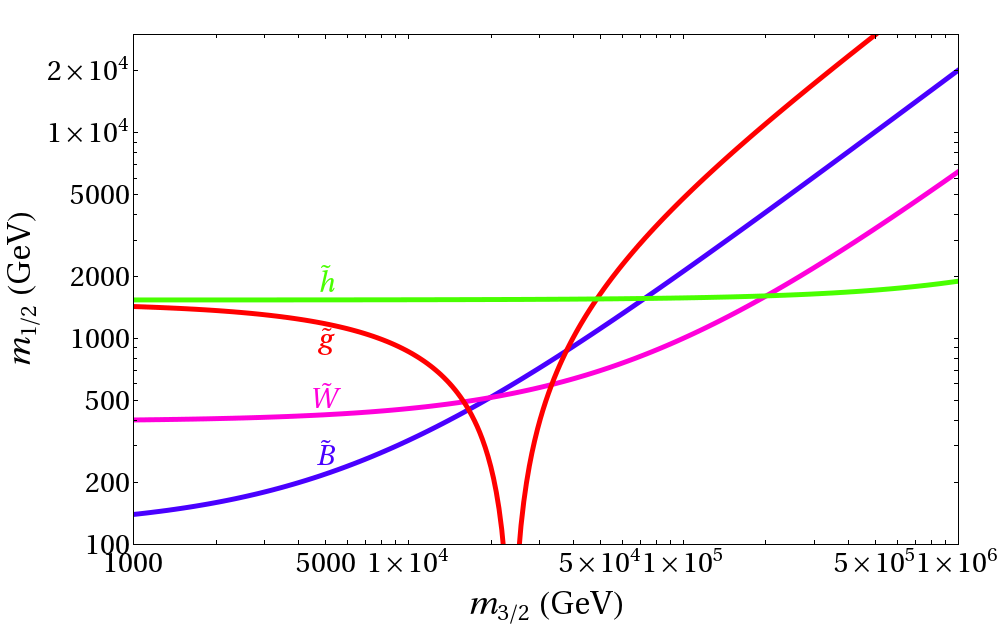}
 \caption{The gaugino and Higgsino mass as a function of the gravitino mass when the SUSY breaking contributions from anomaly and $U(1)'$ mediation are comparable. Here we chose the relative sign of the two contributions so that they add in the bino and wino masses and cancel in the gluino mass. The other choice is also possible and would give a less compressed spectrum.}
 \label{fig:fermionsvsm32}
 \end{center}
\end{figure}

Thanks to the two-loop hierarchy between the scalar and the fermionic spectrum there is an interesting range of values for the SUSY breaking scale ($10^{11}$~GeV~$\lesssim \sqrt{F}\lesssim 10^{12}$~GeV) where the gravitino is heavy enough to alter the gaugino spectrum but light enough not to affect the scalars.
Depending on the gravitino mass the spectrum of the gauginos interpolates between the $U(1)'$ mediated spectrum
(for $m_{3/2}<$~few~TeV) and the anomaly mediated one (for $m_{3/2}>$~few~$10^2$~TeV). In this hybrid mediation case, the hierarchy between the scalars and the gauginos interpolates between two to one loops as the gravitino mass is increased. The flavor-blind contribution of the $U(1)'$ to scalar masses may be used to fix the flavor problem of pure anomaly mediation.

The gaugino spectrum as a function of the gravitino mass is shown in Fig. \ref{fig:fermionsvsm32}. Because of the opposite sign of the $SU(3)$ $\beta$-function compared to the ones for $SU(2)$ and $U(1)_{ Y}$, the anomaly mediated contribution may make the gluino the LSP. As the anomaly mediated contribution increases, the wino now becomes the LSP providing for a calculable DM candidate. The higgsinos can also be light, depending on how EWSB is tuned, allowing for large mixing in the electroweakino sector. This possibility also allows for measuring at colliders the unification of the Split particle couplings to their SUSY values at the high scale.

\subsection{Gauge Mediation}

A Split spectrum can also naturally arise in ordinary gauge mediation. Indeed, it is often a challenge to construct explicit examples of SUSY breaking and gauge mediation that do {\it not} split gauginos and scalars. 

For example, if the theory possesses an $R$-symmetry broken in the hidden sector by a small amount, Majorana gaugino masses are parametrically lighter than the scalars. Alternately, gaugino masses may be suppressed accidentally, i.e., for reasons that are not the result of a symmetry, as in the case of gaugino screening \cite{aglr, cohen}.
Both phenomena are exhibited in a toy model involving the following hidden-messenger interactions:
\begin{equation}
W=M_R\left ( \Phi_1 \bar \Phi_1+ \Phi_2 \bar \Phi_2 \right) + X  \Phi_1 \bar \Phi_2 \,,
\end{equation}
where $\Phi_i, \bar \Phi_i$ are messengers and $X=M+F \theta^2$ is the spurion breaking  SUSY and R-symmetry. When $M < M_R$, the $R$-breaking is small compared to the mass scale of the messengers. The gaugino masses in this case are given by \cite{Buican:2008ws,Cheung:2007es}
\begin{equation}
m_{\lambda_i} = \frac{\alpha_i}{6 \pi} \frac{M}{M_R} \frac{F^3}{M_R^5} +\mathcal{O} \left( \frac{M^3}{M_R^3} \frac{F^3}{M_R^5}, \frac{F^5}{M_R^9} \right)
\end{equation}
while the scalar masses are $\mathcal{O}(\alpha F/M_R)$. Thus the gaugino masses are suppressed relative to scalars by both the smallness of $R$-breaking (via $M/M_R < 1$)  {\it and} gaugino screening (via $F^2/M_R^4 < 1$).  

A related class includes SUSY-breaking models of direct mediation whose low-energy description is a generalized O'Raifeartaigh model with a pseudomoduli space that is locally stable everywhere \cite{Komargodski:2009jf}. This encompasses a wide range of calculable SUSY-breaking sectors, including ISS \cite{Intriligator:2006dd}, the canonical O'Raifeartaigh model \cite{O'Raifeartaigh:1975pr}, and the ITIY model  \cite{Intriligator:1996pu}. In direct mediation models based on such sectors, the mass matrix of fluctuations around the pseudomoduli space has constant determinant, so that the leading-order gaugino mass vanishes. The leading gaugino mass again arises at order $F^3/M^5$, leading to a relative suppression of $\epsilon = F^2/M^4$. 

In all these cases, the higgsinos can be either light or heavy depending on how the EWSB is tuned,
and on the sign of $m_{H_u}^2$, as discussed in sections~\ref{EWSBsection}-\ref{U(1)Split}.
In contrast to ordinary gauge mediation the gravitino can easily be heavier than the gauginos,
and thus allows for the lightest neutralino to be a thermal dark matter candidate.

\subsection{Triplet Mediation}

Another interesting class of Split-type models are those in which one gaugino obtains its mass at a higher loop order than the other gauginos.  Consider, for example, a model where gauge mediation occurs solely through triplet messengers transforming as $(3,0)_{+1/3} \oplus (\bar{3},0)_{-1/3}$ under the SM gauge group. Gauge coupling unification may be preserved if the messenger scale is sufficiently high. The gluino and bino acquire usual GMSB masses at one loop,
\be
m_{\lambda_1} = \frac{2}{5} N \frac{\alpha_1}{4 \pi} \Lambda \hspace{1cm} m_{\lambda_3} = N \frac{\alpha_3}{4 \pi} \Lambda \, ,
\ee
where $N$ is the number of triplet messenger pairs.
Similarly, all scalars acquire GMSB masses at two loops:
\be
m_{\phi_i}^2 = 2 N \Lambda^2 \left[ \left( \frac{\alpha_3}{4 \pi} \right)^2 C_3(i) + \frac{2}{5} \left( \frac{\alpha_1}{4 \pi} \right)^2 C_1(i) \right] .
\ee
However, since there is no $SU(2)$ contribution to SUSY breaking, the smallest soft masses are those for the Higgses and left-handed sleptons, with $m_{\tilde L}^2 = m_{H_{u,d}}^2 = \frac{1}{4} m_{\tilde e}^2.$ Finally, the wino acquires its mass at three loops, of order $m_{\lambda_2} \sim \frac{\alpha_2}{4 \pi} \frac{\alpha_3}{4 \pi} m_{\lambda_3}.$ Thus the wino is separated from the other gauginos and sfermions by two loop factors. In fact, the wino may even be lighter than the gravitino, depending on the separation of the messenger scale and the Planck scale. 

The details of the low-energy spectrum depend upon how $\mu$ and $B_\mu$ are generated. If there are no additional contributions to the Higgs soft masses, then $m_{H_u}^2$ is quickly driven negative and $\mu$ must be large to achieve viable EWSB. In this case, one-loop threshold corrections will raise the the wino mass somewhat, but the wino remains the NLSP (or conceivably the LSP if lighter than the gravitino). In contrast, if $\mu$ and $B_\mu$ are generated at one loop (which requires doublet messengers that do not couple to SUSY breaking, as well as singlet messengers that do), then $\mu$ may be fairly light, between the wino and the remaining sparticles.
This also contributes to the Higgs soft masses, which may render them sufficiently large to avoid tachyonic scalars from RG evolution.  

Similar discussion pertains to models where gauge mediation proceeds through doublet messengers. In this case, the bino and wino obtain masses at one loop; all scalars obtain mass at two loops, with the lightest being the right-handed down-type squarks; and the gluino obtains mass at three loops. Here the problem is that the Higgs soft masses will drive the right-handed stop tachyonic, unless there are additional contributions to the scalar masses.

\section{Mini-Split Phenomenology}
 \label{sec:pheno}
\begin{figure}[t!] 
 \begin{center}
 \includegraphics[width=0.65\textwidth]{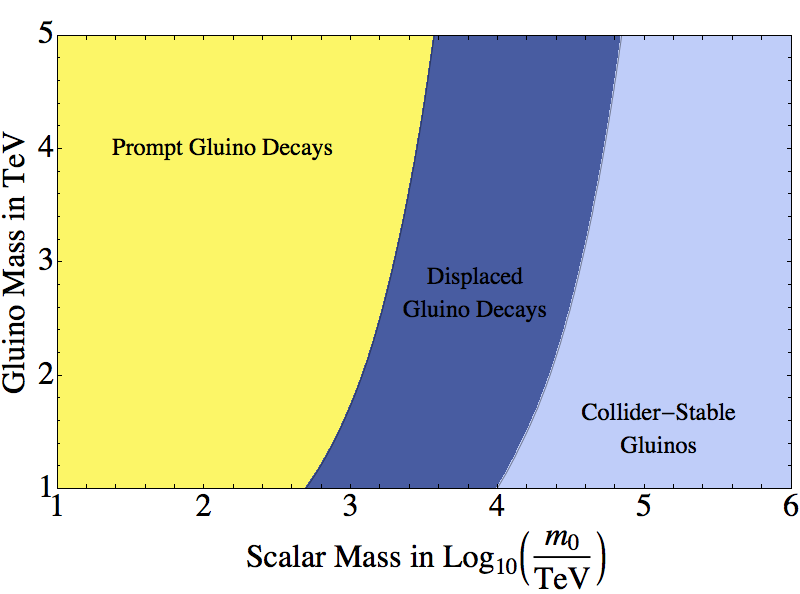}
 \caption{Summary of the gluino decay phenomenology as a function of the gluino mass and the scalar mass scale. We have assumed that for displaced gluinos $100 ~{\rm{\mu m}} \leq c \tau \leq 10~\rm{ m}$.}
 \label{fig:glifetime}
 \end{center}
\end{figure}

If there are no light scalars apart from the Higgs, searching for evidence of supersymmetry at the LHC becomes more challenging but is far from impossible. Of course, the most promising signal is the dimension-6 gluino decay through off-shell scalars to the lightest neutralino state. The lifetime for such a decay can easily vary from a few femtoseconds to Hubble scales as can be seen from Fig.~\ref{fig:glifetime} \cite{glifetime}. The collider signals for such a decay have been studied extensively \cite{Hewett:2004nw, Arvanitaki:2005nq} and there are already bounds for gluinos in Split \cite{Chatrchyan:2012sp, Chatrchyan:2012yg,Aad:2012zn,ATLAS-R} which place their mass above 1~TeV. In addition, as already discussed, taking into account the given Higgs mass and gauge coupling unification, we can infer that the scalar masses in Split scenarios have to lie below $10^5$~TeV. This means that the 
gluino lifetime is now generically less than $\sim 10^{-8}$~sec, and gluinos, when produced at the LHC, will give rise to displaced vertices or prompt decays unless the scalars are above $10^4$ TeV. This makes the search strategies for gluinos in Split more in tune with ordinary gluino SUSY searches.

If the gluino is the ordinary LSP, it instead decays directly to e.g. a gravitino and a gluon. Its decay still gives rise to interesting phenomenology \cite{glifetime}, athough the connection between the gluino lifetime and the scalar masses is lost, and current bounds place its mass above 1 TeV. 

Finally, gluino searches may be supplemented or even supplanted, in cases where the gluino is out of LHC reach, by searches targeting the remaining gauginos or higgsinos. In this case the optimal LHC search strategy depends on the detailed spectrum. The discovery prospects for a light bino with no other accessible states are fairly hopeless, but pure electroweak production of light higgsinos, light winos, or a combination of light states is more promising. 


\subsection{Light Wino and Bino in Mini-Split}

The phenomenology of wino and bino at the LHC greatly depends on the higgsino mass, since the $\mu$ term controls the mixing in the electroweakino sector.

\begin{figure}[t] 
 \begin{center}
 \includegraphics[width=0.8\textwidth]{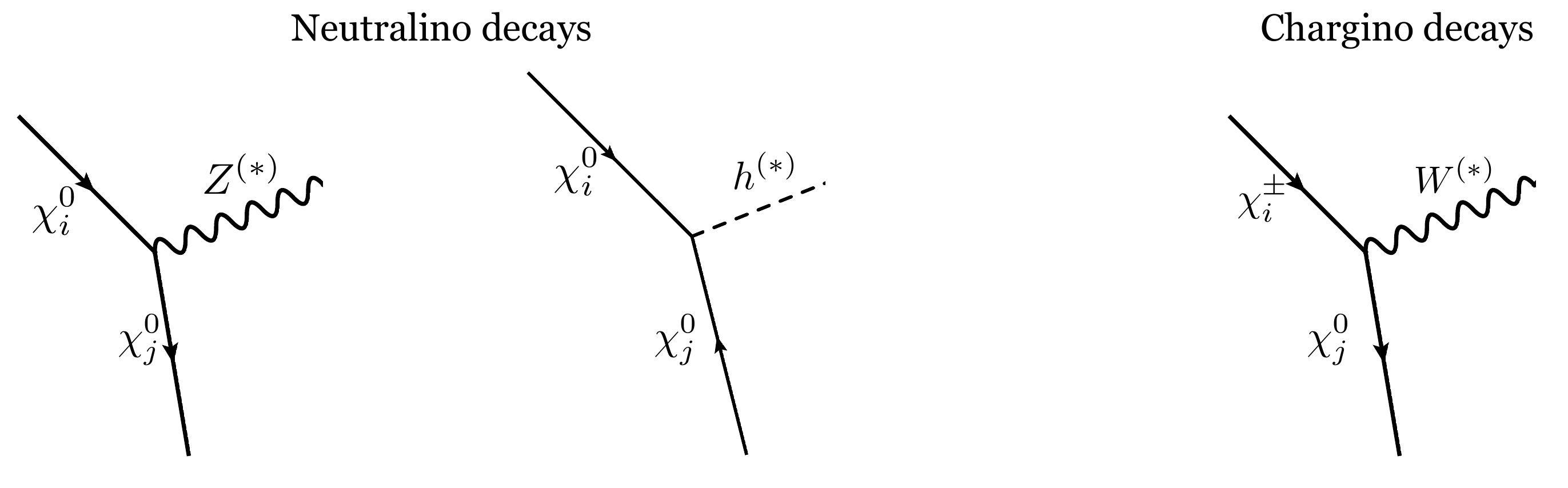}
 \caption{Chargino and neutralino decay channel in Split supersymmetry.}
 \label{fig:decays}
 \end{center}
\end{figure}


\subsubsection{Heavy Higgsinos}

\subsubsection*{$\quad \bullet$ \bf Bino LSP}

If the higgsinos are too heavy to be copiously produced at the LHC, the phenomenology is dominated by the wino and bino and the low-energy spectrum consists of two neutralinos and one chargino. When the bino is lighter than the wino then there may be observable transitions between neutralinos determined by the small higgsino admixture. For small $\tan \beta$, as is most typically the case, the bino-wino mixing angle is proportional to $s_{2 \beta} m_Z^2/ \mu (M_2 - M_1)$. The splitting between the neutral and charged components of the wino is still dominated by Standard Model loops, so the light charginos $\chi_1^\pm$ are nearly degenerate with the wino-like neutralino $\chi_2^0$, with $\Delta m \approx~170$~MeV~\cite{Feng:1999fu}. The dominant production modes are through the wino, via Drell-Yan production of either $\chi_1^+ \chi_1^-$ or $\chi_2^0 \chi_1^\pm$. The $\chi_2^0$ decays directly to $\chi_1^0$ predominantly via $h$ emission, since $Z$ emission is suppressed by an additional power of the small wino-higgsino mixing. 

The chargino branching ratios depend on the size of $\mu$, and for values of $\mu$ consistent with unification typically favor direct decays to $\chi_1^0$ via an on- or off-shell $W^\pm$. Given that the final states are rich in $W$'s, $h$'s, and MET, promising strategies include searches for opposite-sign dileptons plus MET, multi-lepton searches, and searches for the Higgs in association with new physics \cite{Howe:2012xe}.

\subsubsection*{$\quad \bullet$ \bf Wino LSP}

It is often the case that the wino is the ordinary LSP, for example in anomaly-mediated, triplet-dominated, and hybrid $U(1)'$ models. Unless the higgsino is particularly close in mass to the wino, the mass splitting between the neutral and charged wino mass eigenstates is dominated by the electroweak contribution, $\sim 170$ MeV for an electroweak triplet (and roughly half the size of the splitting for electroweak doublets). This implies that decays of $\chi^\pm$ to the neutral component proceed much as in the minimal model, predominantly to soft charged pions. However, the smaller splitting in this case leads to a significantly greater lifetime, corresponding to $c \tau \sim 10$ cm at the LHC. This increased length is comparable to the scale of the central trackers at ATLAS and CMS, so that the charged stubs are conceivably observable. The optimal search strategy entails studying processes with at least one ISR jet and triggering on MET or jet $p_T$, followed by offline analysis to identify the charged stubs \cite{Buckley:2009kv}; such searches are already underway at the LHC \cite{ATLAS:2012ab}.

\subsubsection{Light Higgsinos}

If the higgsino is light, the range of LHC signals grows richer still. Now there may be appreciable mixing between the higgsino, bino, and wino, and cascades among neutralinos and charginos are prompt. The most promising signals are again the pair production of charginos and/or neutralinos, followed by cascades to the lightest neutralino via $W$'s and $h$'s. In addition, the presence of light higgsinos raise the prospects of cascade decays between neutralinos involving $Z$'s.  The full set of decay channels is shown in Fig.~\ref{fig:decays}.

If all neutralinos and charginos are accessible, a measurement of their couplings at a future linear collider can also provide for another measurement of the scalar mass scale~\cite{ad04}. If we run the measured couplings to the UV from the low scale, we should find that they unify to their SUSY values at the same scale. Measuring their couplings with 10\% accuracy should allow the determination of the scalar masses within a few orders of magnitude. If the gluino is kinematically accessible, this measurement may be cross-checked against the scalar mass inferred from gluino decays, potentially allowing the determination of $\tan \beta$.

\subsection{Higgsino LSP: \newline The minimal model for Unification and \mbox{Dark~Matter}}

It has been already pointed out \cite{kachru, senatore} that the minimal realization of Split Supersymmetry that preserves unification and has a calculable dark matter candidate is the Standard Model plus a pair of light higgsinos. Even a tiny admixture of bino is enough to evade bounds on Dirac Dark Matter, which can easily happen as long as the bino is lighter than $\sim10^3-10^4$ TeV. As discussed earlier, such a scenario may arise in, e.g., anomaly-mediated mini-Split if the gauginos are somewhat heavier than the higgsinos. 

The primary challenge for the minimal model is the discovery of the higgsinos. In the limit where only the higgsinos are light, the mass splitting between the charged and neutral mass eigenstates comes from one-loop corrections involving photons and $Z$'s \cite{Dimopoulos:1990kc, Thomas:1998wy}. While the splitting is a function of the ratio $\mu^2 / m_Z^2$, for $\mu^2 \gg m_Z^2$ it saturates at $\delta m = \frac{1}{2} \alpha m_Z \simeq 355$ MeV. Thus chargino pairs produced via Drell-Yan processes will decay to their neutral partners with relatively soft associated decay products. 

For splittings on the order of $\sim 355$ MeV, the available decay channels are $\chi^\pm \to \chi^0 \pi^\pm, \chi^0 \ell^\pm \nu$ with $\ell = e, \mu$. Decays to pions dominate by two orders of magnitude, largely due to the effects of two-body vs. three-body phase space, though in both cases the decay products are soft, and even soft leptons would be challenging to identify. The value of $c \tau$ for these decays is slightly less than 1 cm, so that the charged tracks from $\chi^\pm$ are not visible at the LHC, while the pion is too soft to be efficiently distinguished from backgrounds. One might hope to search for $\chi^+ \chi^-$ pair production at the LHC in association with an ISR jet or a pair of jets from vector boson fusion, but the backgrounds for these processes are prohibitively high if there is no information about the charged stub. However, given sufficiently high integrated luminosity, it may be possible to trigger on longer charged tracks in the exponential tail of chargino decays \cite{Buckley:2009kv}; this possibility warrants further detailed experimental study at the LHC.

  Alternately, at a future $e^+ e^-$ collider the optimal search strategy involves triggering on a hard radiated photon from ISR and then looking for the soft pions \cite{Thomas:1998wy}. The dominant background is $\nu \nu \gamma$, which is nearly two orders of magnitude larger than the typical cross section for $\chi^+ \chi^- \gamma$ production, so information about the soft pions or charged stubs would still be required. Nonetheless, it is possible that clever experimental search strategies may improve the prospects for discovery.

The situation improves in models with low-scale SUSY breaking where the gravitino is the LSP and the higgsino decays inside the detector. In this case the much easier channel $hh$+MET is available and higgsinos lighter than a few hundred GeV should become accessible to the LHC after the upgrade.

\subsection{The Higgs in Mini-Split}
\begin{figure}[t!] 
 \begin{center}
 \includegraphics[width=\textwidth]{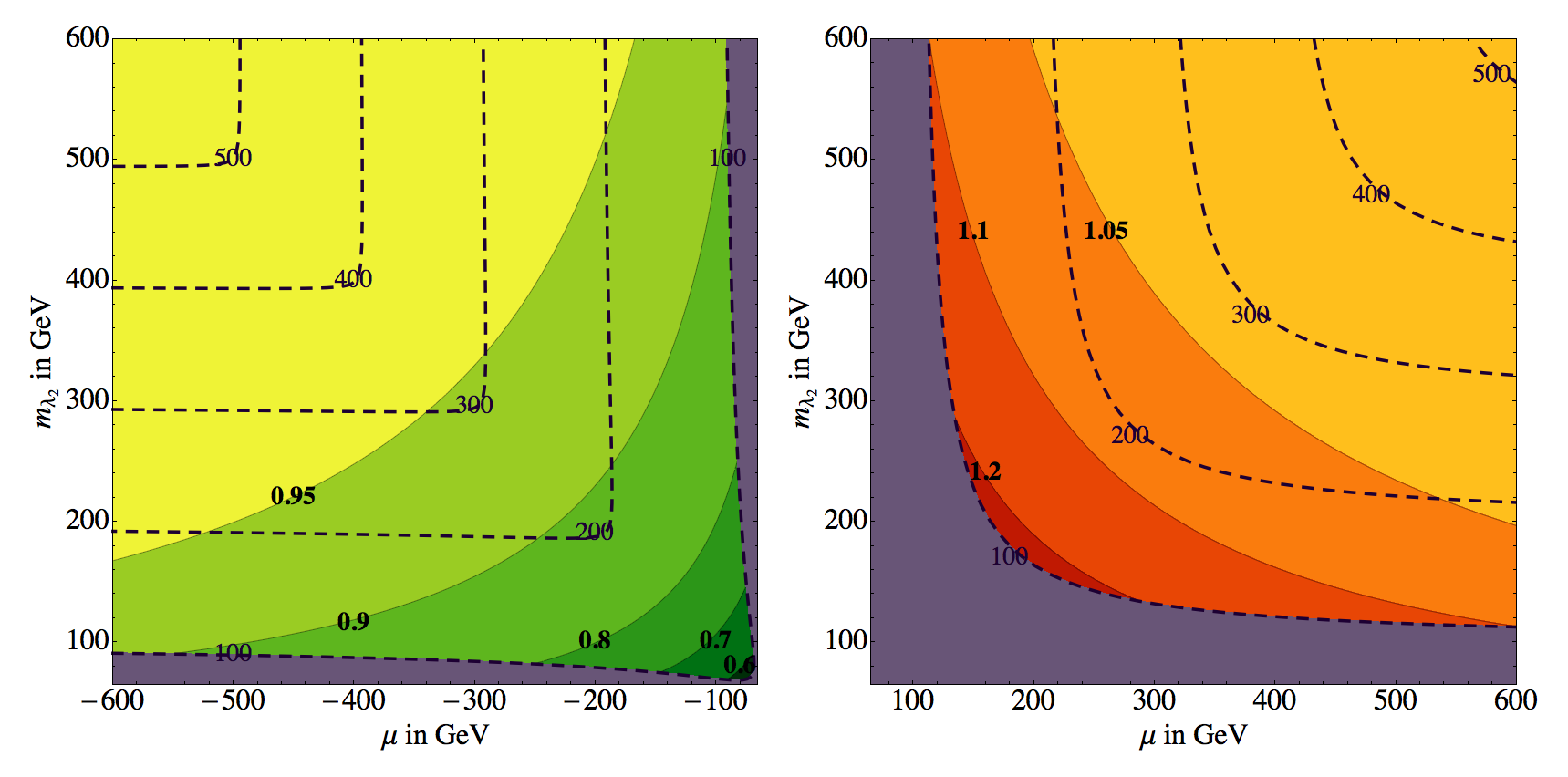}
 \caption{Contours of $\Gamma_{h\to\gamma\gamma}/\Gamma^{SM}_{h\to\gamma\gamma}$ in the higgsino-wino mass plane for $\mu m_{\lambda_2} < 0$ (left) and $\mu m_{\lambda_2} > 0$ (right) with $\tan \beta = 1$. The dashed contours denote the lightest chargino mass in GeV.  The purple-shaded region indicates the LEP2 exclusion of charginos lighter than $\sim$ 100~GeV.}
 \label{fig:higgs-chargino}
 \end{center}
\end{figure}

In natural models of EWSB the Higgs boson is not a SM Higgs, as the new physics solving the hierarchy problem sensibly changes the Higgs properties. In supersymmetry, the leading quadratic divergences are cancelled by the stop, which also alters the Higgs couplings to gluons and photons. The absence of new physics at the LHC so far suggests
that the Higgs properties are unlikely to deviate significantly from the SM. On the other hand, in Split supersymmetry the electroweak scale is tuned, the scalars are heavy, and their effects on the Higgs properties decouple. In this case we expect the Higgs to be completely SM-like. The only exception is if the charginos are light, in which case they may contribute to the effective coupling between the Higgs and photons. The effect of light charginos on the branching ratio of $h\to\gamma\gamma$ as a function of $\mu$ and $m_{\lambda_2}$ is illustrated in Fig.~\ref{fig:higgs-chargino}. 
The qualitative features of their contribution are captured by the approximate formula
\begin{equation}
\frac{\Gamma_{h\to\gamma\gamma}}{\Gamma^{SM}_{h\to\gamma\gamma}}
\simeq 1+\frac{12}{17}\frac{m_W^2 \sin2\beta}{\mu\, m_{\lambda_2} -m_W^2 \sin2\beta}\,,
\end{equation}
valid in the decoupling limit for the charginos.  The Higgs branching ratio to $\gamma\gamma$ is enhanced (suppressed)
compared to the SM when $\mu\, m_{\lambda_2}$ is positive (negative). Since the Higgs couples to charginos only through their mixing, the effect is maximized when both charginos are light and $\tan\beta=1$. The deviation from the SM result is between $-40\%$ and $+20\%$ once the bounds  on the lightest chargino from LEP2 are taken into account.

Thus the properties of the Higgs in Split SUSY are the same as in the SM, unless both charginos are very light and $\tan\beta$ is close to unity. In this case only the branching ratio to $\gamma\gamma$ is significantly affected. If only the Higgs coupling
to photons is found to differ from its SM value, Split SUSY models predict charginos within reach at the LHC.

\section{Conclusions}

The continued absence of any new particles at the LHC diminishes the connection between the electroweak scale and new physics and points to a fine-tuned theory. The only evidence we have for low-scale supersymmetry is gauge coupling unification. This, together with the measured higgs mass, makes a case for Split SUSY with scalars below $10^5$~TeV. If the flavor problem is not addressed, FCNCs force the scalars to be above a few thousand TeV. At the same time, unification requires the higgsinos to be light, in which case EWSB occurs
only when $m_{H_u}^2>0$. Therefore in any Split model either the flavor problem or the tachyon problem must be solved.

Gauginos and higgsinos may be within the reach of the LHC, giving rise to displaced gluino signatures as well as cascade decays between the neutralinos and charginos through W, Z, and higgs emission. \href{http://www.youtube.com/watch?v=IhJQp-q1Y1s}{Nevertheless, nature leaves open the daunting possibility that all sparticles are in the multi-TeV range away from LHC reach while still in agreement with the higgs mass and unification.} On the other hand, if gauginos are observed at the LHC, a detailed study of their properties at a linear collider may be critical to confirm their supersymmetric origin and measure the SUSY breaking scale.

\section*{Acknowledgements}

We thank Nima Arkani-Hamed, Masha Baryakhtar, Timothy Cohen, Sergei Dubovsky, Tony Gherghetta, Kiel Howe, David E. Kaplan, John March-Russell, Jesse Thaler, and Yue Zhao for interesting discussions, and Gino Isidori and Alessandro Strumia  for clarifications on the computations of ref. \cite{strumia}. We would also like to thank the theory group at CERN for  hospitality and A~Nosa~Galiza for sustenance during the completion of this work.

This work was partially supported by ERC grant BSMOXFORD no. 228169. NC was supported in part by DOE grant DE-FG02-96ER40959 and the Institute for Advanced Study.

\end{document}